# Analysis of First Order Reversal Curves in the Thermal Hysteresis of Spin-crossover Nanoparticles within the Mechanoelastic Model

Laurentiu Stoleriu[1*], Alexandru Stancu[1], Pradip Chakraborty[2,†], Andreas Hauser[2], Cristian Enachescu[1*]

[1]*Faculty of Physics, Alexandru Ioan Cuza University, 700506, Iasi, Romania*

[2]*Département de Chimie Physique, Université de Genève, CH-1211, Switzerland*

*Corresponding authors: lstoler@uaic.ro, cristian.enachescu@uaic.ro.

*

**Abstract**

The recently obtained spin-crossover nanoparticles are possible candidates for applications in the recording media industry as materials for data storage, or as pressure and temperature sensors. For these applications the intermolecular interactions and interactions between spin-crossover nanoparticles are extremely important, as they may be essential factors in triggering the transition between the two stable phases: the high-spin and low-spin ones. In order to find correlations between the distributions in size and interactions and the transition temperatures distribution, we apply the FORC (First Order Reversal Curves) method, using simulations based on a mechanoelastic model applied to 2D triangular lattices composed of molecules linked by springs and embedded in a surfactant. We consider two Gaussian distributions: one of the size of the nanoparticles and one of the elastic interactions between edge spin-crossover molecules and the surfactant molecules. In order to disentangle the kinetic and non-kinetic parts of the FORC distributions, we compare the results obtained for different temperature sweeping rates. We also show that the presence of few larger particles in a distribution centered around much smaller particles dramatically increases the hysteresis width.

**Introduction**

The study of spin-crossover nanoparticle systems has become a hot topic in the spin-crossover community over last few years.[1-3] Several groups have synthetized spin-crossover nanoparticles and have investigated their properties for possible applications as sensors or for data storage.[4] Consequently, several unexpected differences between nanoparticles and bulk compounds have been detected. The thermal transition from the high-spin state (HS, stable at high temperatures) towards the low-spin state (LS, stable at low temperature) is usually shifted towards lower temperatures and the accompanying hysteresis, if it is still observed, has a smaller width.[3] In addition, the transition is smoother and sometimes incomplete, producing HS residual fractions, which persist at lower temperatures. This behavior has been interpreted in terms of a reduced cooperativity, as the use of open boundary conditions limits the effect of the long-range interactions.[5] In addition, comparing to bulk systems, the thermal transition in nanoparticle systems can be strongly affected by kinetic[6] or matrix[2,7] effects.

The elaboration of models aiming to explain the behavior of spin-crossover nanoparticles appeared as a logical consequence of their synthesis. An initial Ising-like model has been elaborated by Kawamoto and Abe[8] a couple of years before the production of first experimental nanoparticles. This model considered only the short range interactions and could explain only partially the behavior of nanoparticles (the reducing of the hysteresis width and the smaller cooperativity). The Ising-like model has been recently improved in order to explain most of spin-crossover nanoparticles behavior, either by considering the edge molecules constrained to the HS state[9] or taking into account the interactions with neighboring molecules of surfactant[10]. Intending to remove some drawbacks of the Ising-like model (like the impossibility to predict the real evolution of clusters as reflected in optical microscopy experiments[11]), the elastic models take into account the molecular volume change during the HS-LS transition and, consequently, the elastic interactions in this model appear as a direct result of the spatial distortions inside the sample. Among several elastic models, the mechanoelastic model has been successfully adapted for characterizing the thermal hysteresis either in spin-crossover bulk[12], microparticles[13] or nanoparticles[12], by considering elastic interactions between the crystal and the environment, which may be described as a variable external pressure.[7] Recently, the size-dependent rigidity of nanoparticle was also explored in simulations as a possible cause for some experimental results on spin-crossover nanoparticle systems.[14]

A sensitive method to characterize the hysteresis properties in spin-crossover materials is the First Order Reversal Curves method. This method is often seen as a general, model independent technique which gives a distribution that can be linked to the interaction and domain size distribution in the system and, therefore, can be applied virtually to any type of hysteretic process (ferromagnetic, ferroelectric, thermal, elastic, geological etc.)[15]. In spin-crossover compounds, the FORC method has been applied for the characterization of the so-called like spin models, during the thermal[16] or the pressure[17] hysteresis and for a fine tuning of interactions in the case of compounds doped with impurities.[18] In recent experimental work, Varret et al have shown FORCs for spin-crossover nanoparticles, showing a large reversible component accompanied by a strong kinetic effect.[6] A similar behavior has been reported for $Fe(phen)_2(NCS)_2$ spin-crossover microparticles embedded in various surfactants.[7]

In this paper we simulate FORCs for spin-crossover compounds in the framework of the mechanoelastic model[12] and discuss them in terms of size and the strength of the elastic interactions. Even if the mechanoelastic model can reproduce qualitatively most of the special features of the spin-crossover nanoparticle hysteretic behavior, it cannot be used to calculate minor hysteresis loops (inside the hysteresis loop) as for a single particle it allows only the existence of two stable states. In order to overcome this inconvenience, we consider here distributions of nanoparticle sizes and interactions with the environment and the states of the whole system are calculated as a superposition of individual states.

The paper is organized as follows: first we present the mechanoelastic model and then introduce the Gaussian distributions of sizes and interactions with the environment. After discussing the role of the width of the distribution on the hysteresis properties, we compare FORCs for different kinetics to discriminate between kinetic and static parts of the FORC diagrams.

**Results and discussions**

The mechanoelastic model (ME) is based on the so-called ball and spring concept[19] and implies that the elastic interactions inside the sample arise from lattice distortion due to the difference of the molecular sizes between the LS and HS states. The spin-crossover molecules are represented as small spheres, situated in triangular open boundary lattices and interacting by springs, with the elastic constant $k_{el}$. The volume change of a switching molecule produces an instantaneous elastic force in its neighboring springs and consequently determines the position shift first of its closest neighbors and then of all other molecules in the system.

The non-periodic conditions intrinsic to the ME model make it appropriate for the study of behavior of spin-crossover nanoparticles. An initial theoretical study on the thermal hysteresis of spin-crossover entities of different sizes using this model has reproduced the main experimental features described in the Introduction.[12] In a first approach, taking into account only the system size variation, only the smoother transition and the smaller hysteresis width could be reproduced. Therefore, in order to account for the other experimental features (shift of the transition towards lower temperatures with residual HS fractions) we have considered the spin-crossover samples embedded in a surfactant surrounding environment. For practical reasons, the polymer has been represented as a rigid shell of non-switching molecules which interact with the molecules situated at the edge of the lattice by way of springs, with a given elastic constant $k_{poly}$. In this way, the reduction of the overall volume of the nanoparticles while the transition proceeds induces an increasing negative pressure at the edge of the system, originating in pulling forces from the polymer network.[10, 12, 13]

In order to study the thermal transition within the ME model, one assigns for every spin-crossover molecule a HS→LS or a LS→HS transition probability, depending on the temperature $T$, on intrinsic material parameters (the HS-LS energy difference $D$, the degeneracy ratio $g$, the effective activation energy $E_A$) and on the interactions between molecules, represented here by the way of a local pressure $p_i$, defined as the sum of elastic forces acting on the molecule and taken positive for compressed and negative for elongated springs, according to:

$$P^i_{HS \to LS} = \frac{1}{\tau} \exp\left(-\frac{E_A - \kappa p_i}{k_B T}\right)$$

$$P^i_{LS \to HS} = \frac{1}{\tau} \exp\left(-\frac{D - k_B T \ln g}{k_B T}\right) \exp\left(-\frac{E_A + \kappa p_i}{k_B T}\right),$$

where $i$ runs over all molecules in the system, $\tau$ is a constant factor chosen so the probabilities are well below unity at every temperature and $\kappa$ a scaling factor between the local pressure and its effect on the activation energy of the individual molecule.[12] Following a Monte-Carlo standard procedure, one decides whether a molecule switches or no. A Monte-Carlo step (MCS) is completed when every molecule has been checked once. After every MCS, all molecules are allowed to move in order to find their new mechanical equilibrium positions, implying changes in local interaction values. In the simulations presented here the following parameters have been used: $\tau = 100$, $E_A = 400 K$, $k_{el} = 1 N/m$ $\kappa = 2000 K/N$, $k_B = 1$, $D = 1100 K$, $\ln g = 5.5$.

As stated above, this simple model is not able to reproduce minor hysteresis curves for a single particle. Previously, the minor hysteresis loops have been simulated either by directly considering distributions of switching temperatures, for instance in Preisach-type models[20, 21], or implicitly, by considering the distributions of intrinsic parameters, like the activation energy[22]. In this work, we consider a more realistic approach by taking into account a distribution of nanoparticles sizes, similar to that experimentally observed[23]. In addition to this size distribution, we consider a distribution of the constant designating the interactions with neighboring polymer matrix.

In order to generate these distributions, we have considered systems of up to $10^8$ isolated nanoparticles embedded in polymer. Discrete Gaussian distributions of nanocrystallite sizes - (number of particles per hexagonal side) from 5 to 30 particles per side corresponding to hexagons with 61 to 2611 particles and interactions with the polymer (7 different values of $k_{poly} = 0.4...1.0 \times k_{el}$ ) have been taken into account for most of the simulations presented in this paper. Every nanoparticle in the system responds somewhat differently to the temperature variation, according to its size and the strength of interaction with the environment. In Fig.1, we represent a snapshot of different sizes spin-crossover nanoparticles embedded in a polymer environment, together with the distribution of sizes and intermolecular interactions.

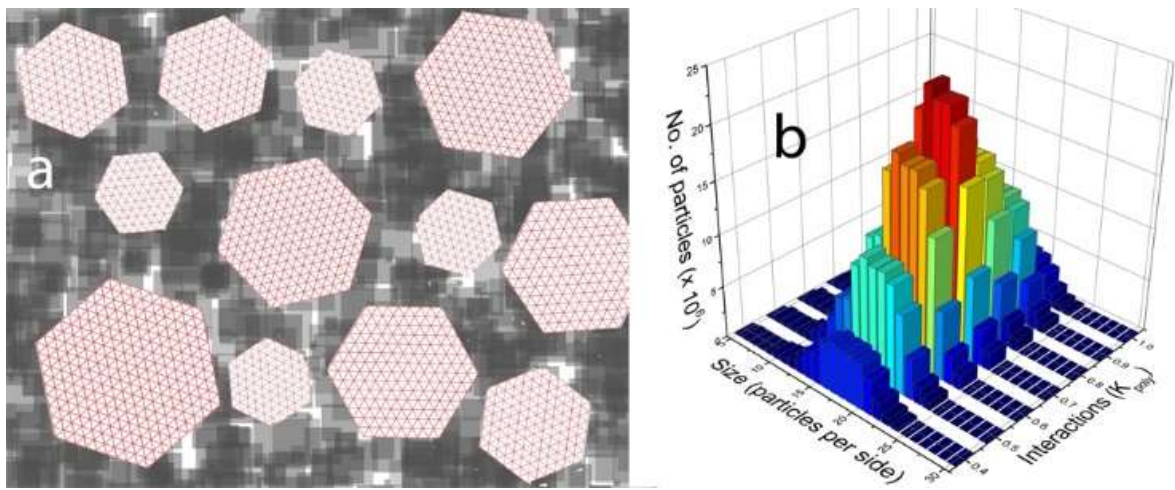

**Figure 1 (a) Example of system composed of different size hexagonal nanoparticles in a polymer environment (b) Gaussian size and interaction distributions**

We have simulated separately the transition curves for every kind of nanoparticle. For each combination of *size* and *interaction* parameters, up to 100 runs (depending on the hexagon sizes) have been made and for two temperature sweeping rates: 100 MCS/K and 500 MCS/K. One run contains the major hysteresis loop as well as the first-order reversal curves. The final curves considered for each combination of parameters are obtained by averaging of all the runs.

In Fig. 2 we present the samples size and interaction with the influence of the environment on the hysteresis loop: generally, the smaller the size, the larger the shift towards lower temperatures and the narrower the hysteresis. The same effect is obtained by increasing the interaction with the polymer. We notice that for some systems, especially those with smaller sizes and strong interactions, a residual HS fraction is observed. Based on the size and interactions distribution one can calculate the distributions of switching temperatures, defined as the temperatures at which the fraction of the HS molecules in the system, denoted here as $n_{HS}$ is equal to 0.5; these distributions are represented in the inset of Fig.2 and are, in a good approximation, Gaussian.

Once all the curves for all the combinations of parameters are obtained, one can use simple algebraic summing of elementary loops to build the total curves of the system based on custom distributions of parameters.

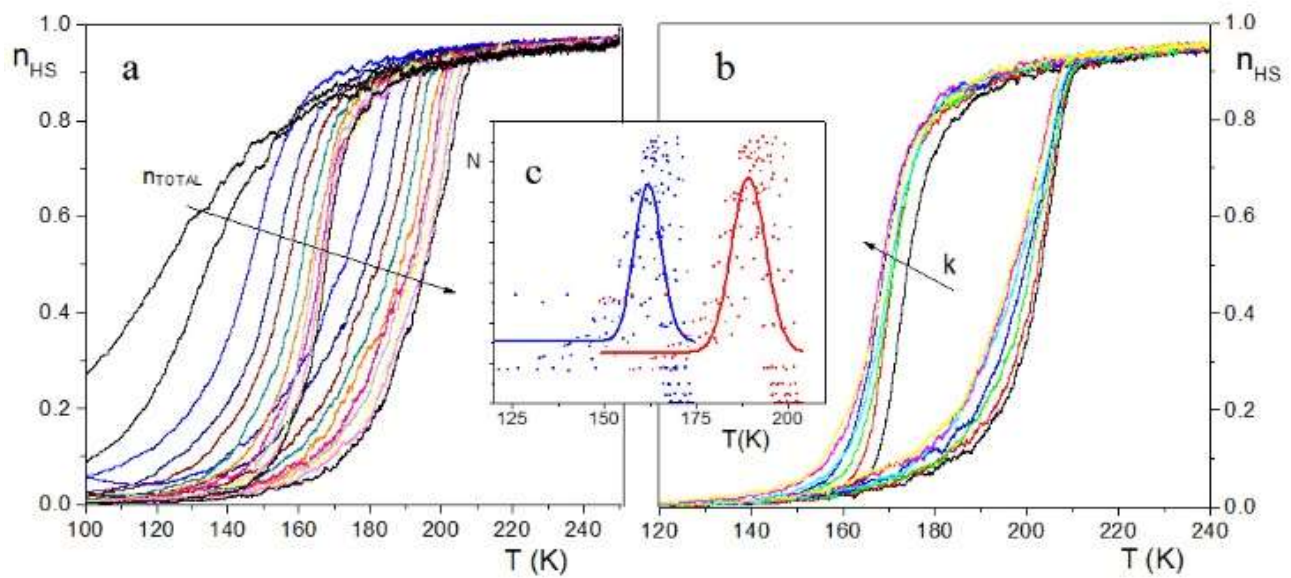

**Figure 2. (a) Thermal hysteresis for various nanoparticles sizes, when interactions with the environment ($k_{poly} = 0.7 \times k_{el}$) are kept constant. (b) Thermal hysteresis for various**

**interactions with environment for the same nanoparticle size (15 molecules on every side), (c) Switching temperatures distributions for the distribution presented in Fig 1b**

Before calculating the FORCs and analyzing the FORC distributions, let us analyze the effect of the size distribution in a simpler situation corresponding to the absence of interactions with the environment. In Figure 3c we present the thermal transition curves for two systems with Gaussian size distributions centered at the same particles diameter and with the same width, when one of these systems contains also some larger particles, in a small quantity (1% of the total). These larger particles are practically not visible on Figure 3a, where both distributions are represented in terms of number of systems corresponding to each size. However, as the effect of one large particle is much bigger than the effect of a smaller one due to its higher contribution to the magnetic signal, the distribution represented as the number of molecules corresponding to systems sizes is quite different in the two cases (Figure 3b). As smaller nanoparticles show a smooth transition with no hysteresis and large nanoparticles present a hysteresis, the transition curves are very different in the two cases. Even if the large majority of the particles in the second system do not show an individual hysteresis, the whole system exhibit hysteresis due to the effect of larger particles. This conclusion is important for experimental situations: the experimental data for nanoparticle systems with size distributions centered at a few nanometers in diameter and showing a hysteresis should be carefully treated as they might be "contaminated" by a very small number of larger nanoparticles.

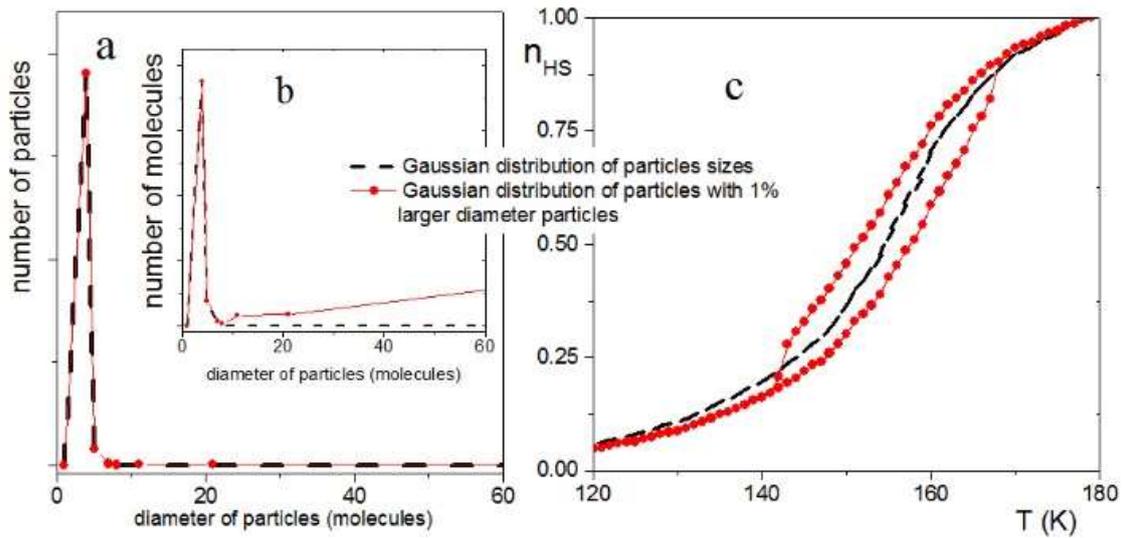

**Figure 3 Effect of the width of the distribution (a, b as described in text) on the hysteresis cycle (c) (in the case with no interaction with the polymer)**

First order reversal curves are obtained by reversing the variation of the input parameter, in our case temperature, while the transition is still in development. The FORC diagrams are contour plots of the second order mixed derivative of the HS fraction with respect to the sweeping temperature $T$ and the reversal temperature $T_r$:

$$\rho(T, T_R) = -\frac{\partial^2 n_{HS}}{\partial T \partial T_R}$$

Where $\rho(T, T_R)$ is the FORC distribution.

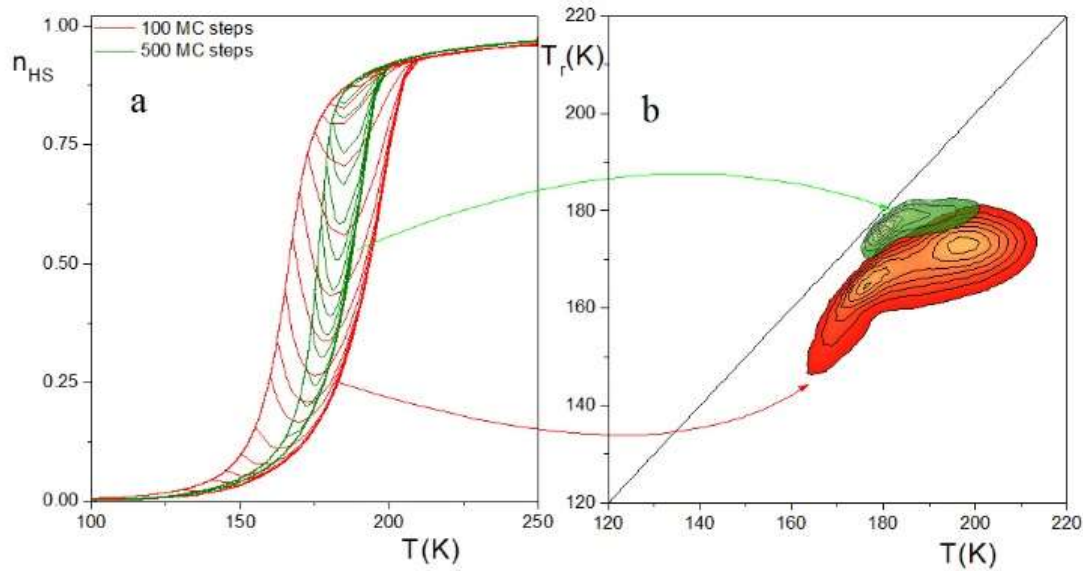

**Figure 4 Hysteresis loops and FORC diagrams for two temperature sweep rates, for a distribution of particles as presented in Fig. 1**

In general, the hysteresis loops corresponding to spin-crossover nanoparticles, similar to nanoparticles of other magnetic materials, present strong kinetic effects [24, 25]. Recent experimental measurements[6] have shown that sizable kinetic effects are accumulated mostly in the proximity of the major hysteresis loop. Similar kinetic effects are reproduced in simulations based on Monte Carlo methods using either Metropolis or Arrhenius dynamics, due to the finite computing time available for finding the exact HS fraction at a given temperature. Therefore, in order to disentangle the kinetic behavior reflected in the FORC distributions from the static behavior, we have simulated FORCs for two temperature sweeping rates (Fig. 4a). As demonstrated by this figure, the number of MC steps per unit temperature dramatically influences the shape and the width of the thermal hysteresis: a fast temperature sweep rate results in an extra broadening of the thermal hysteresis.

Fig.4b displays the FORC diagrams of the two systems discussed above, showing a kinetic distribution along the first bisector and a central distribution corresponding to the irreversible component. The kinetic distribution can be obtained even in the absence of any parameter distribution, if the sweep rate is quite large, while the central distribution tends to the static distribution for very low temperature rates.

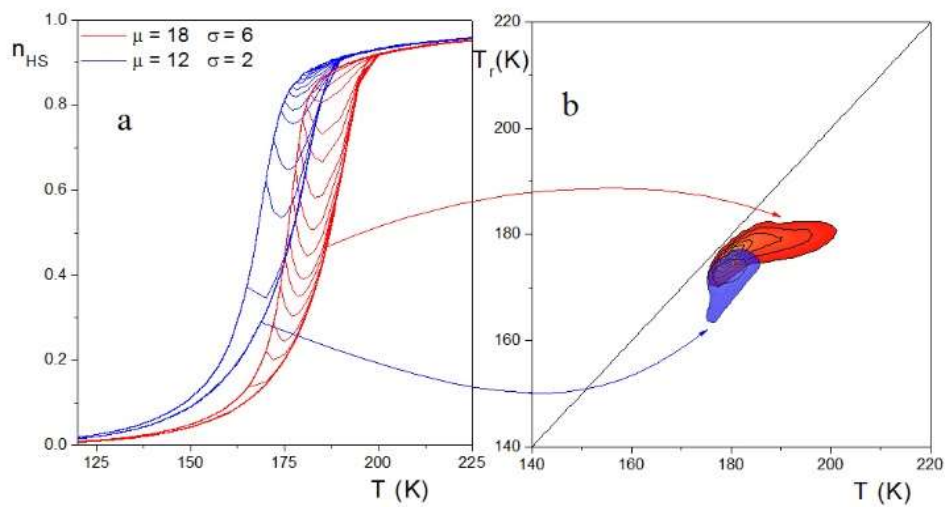

**Figure 5 Hysteresis loops and FORC diagrams for two distributions, for the same temperature variation rate (500 MC steps/K). Sizes distribution (center μ and width σ) are mentioned in the legend**

In Fig.5 there are represented FORCs and their specific diagrams for two nanoparticle systems with different Gaussian size distributions and embedded into the same surfactant (which results in the same interaction distribution) and with the same temperature sweep rate (500 MC steps/ K). In the first case (blue curve) the irreversible component is smaller due to the smaller cooperativity characteristic for smaller size particles. For systems with higher cooperativity, as in the second diagram (red curves), the irreversible component is better distinguished. It can be observed that for the first system in which the cooperativity is weaker, the widths of the hysteresis loop and quasistatic hysteresis loop are both smaller. Consequently, the system is characterized mainly by kinetic effects. However for the second system in which the cooperativity effect is substantial, both the widths of major hysteresis loop and quasistatic hysteresis loop are wider. The kinetic component of the system can be clearly distinguished only at the boundaries of MHL, outside the quasistatic hysteresis loop.

**Conclusions:**

In the present work we have proposed a theoretical analysis based on the FORC diagram method for thermal transitions of systems of spin-crossover nanoparticles with size and interaction distributions, based on a the mechanoelastic model coupled to a Monte Carlo algorithm with Arrhenius dynamics. The interaction between the molecules of the nanoparticles and a polymer matrix in which the nanoparticles are synthesized was also considered. The FORC curves and diagrams are compatible with available experimental data and can provide a method to determine the distribution of nanoparticle sizes and evaluate the interactions inside the samples.

## Acknowledgements

This work was supported by Romanian CNCS – UEFISCDI, project number PN-II-RU-TE-2011-3-0211 and by Swiss National Science Foundation (Grant No. 200020-137567)

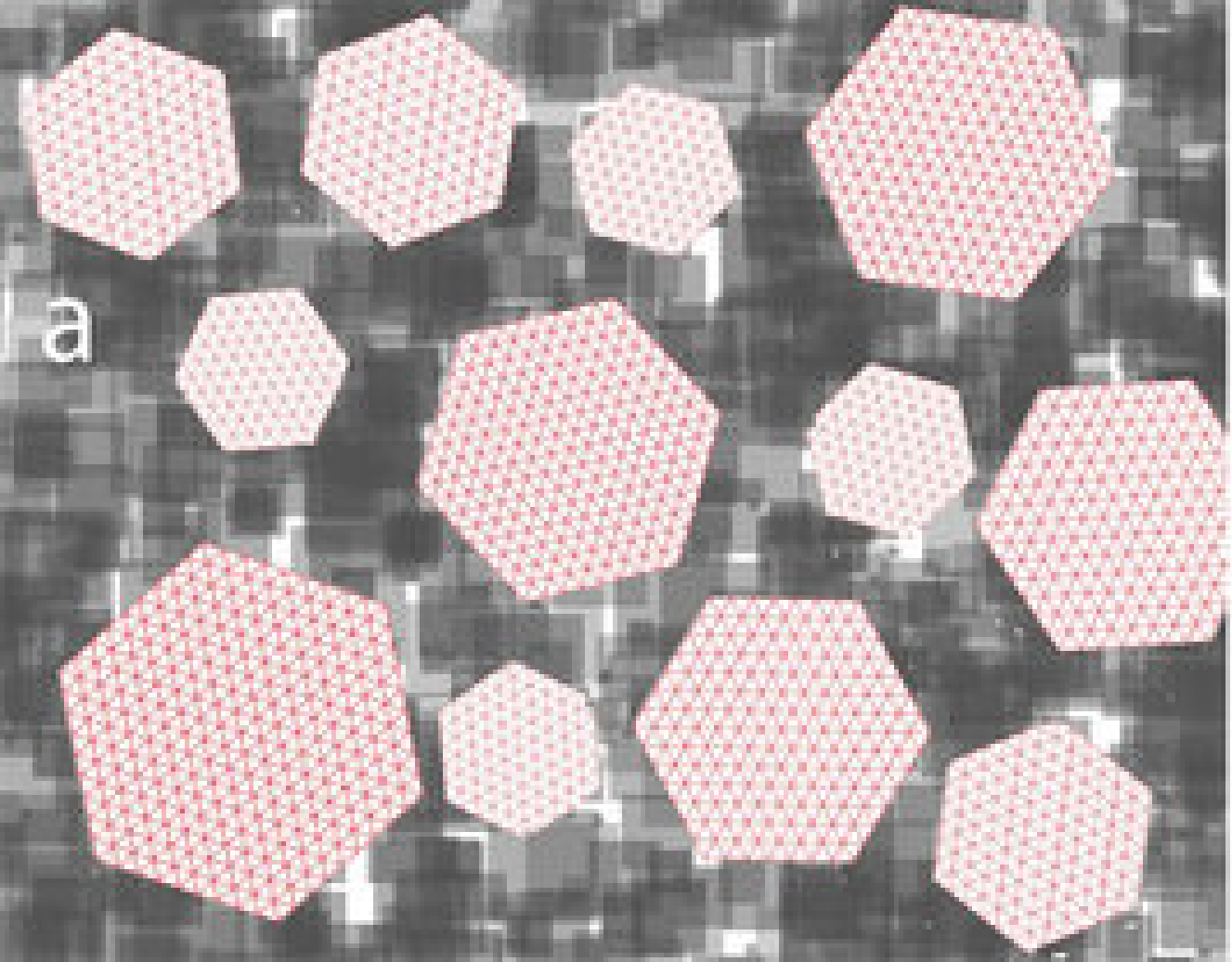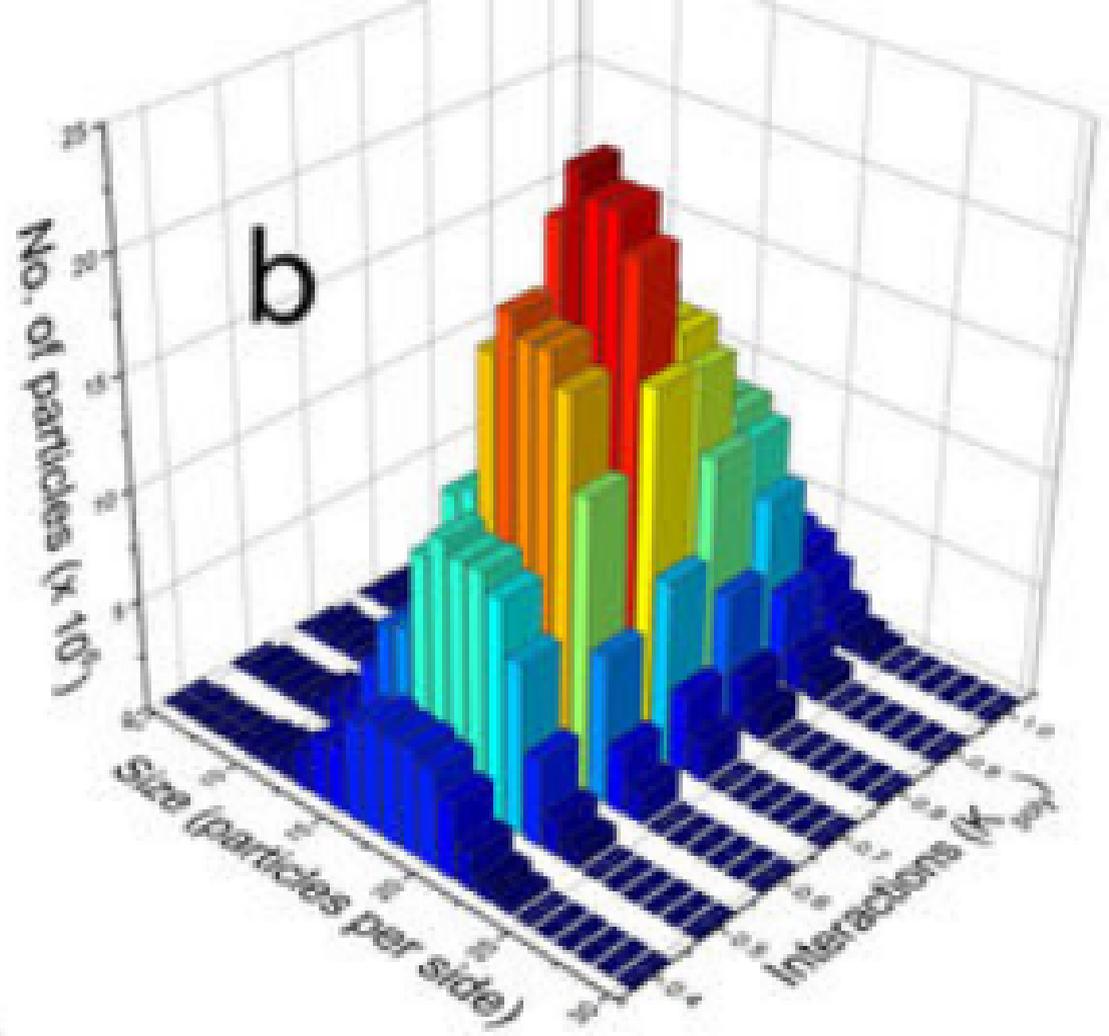

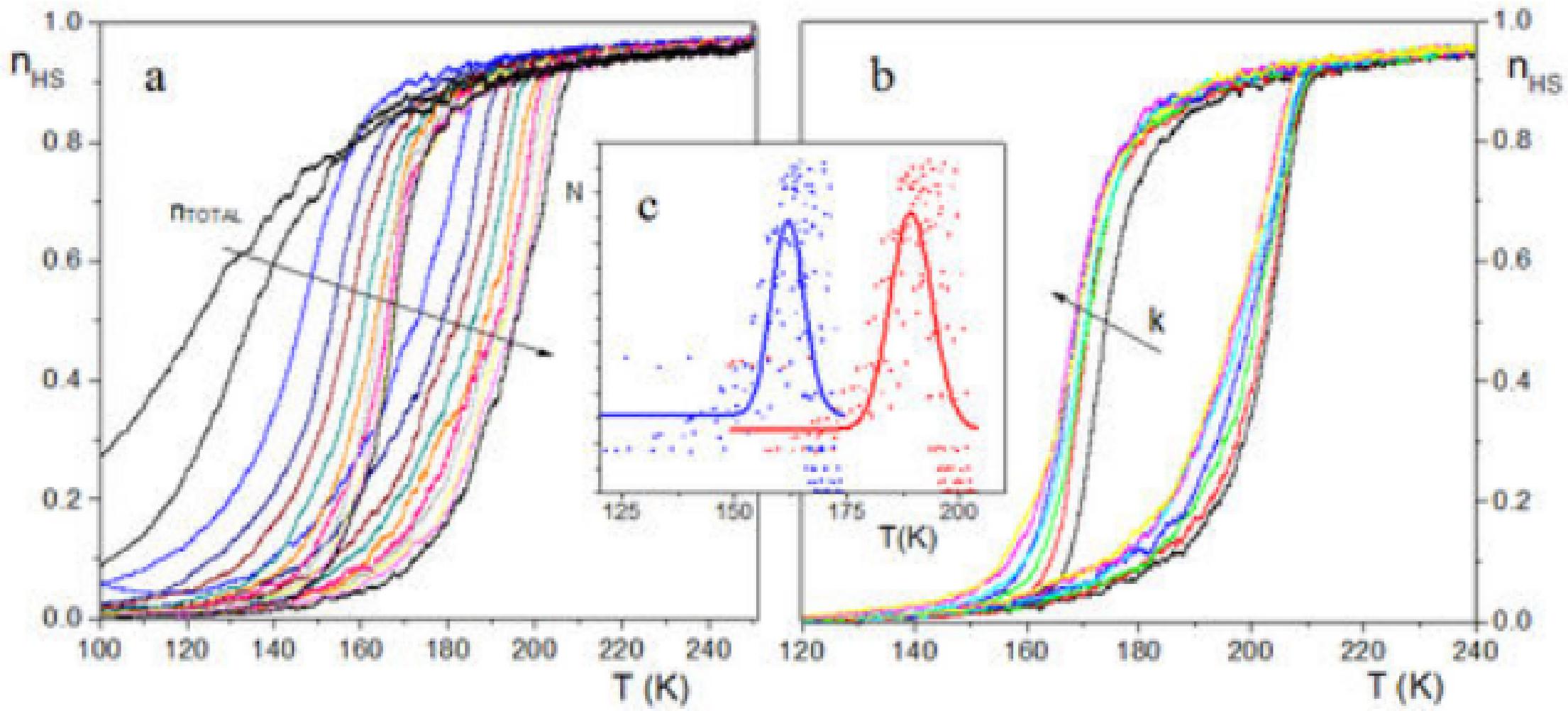

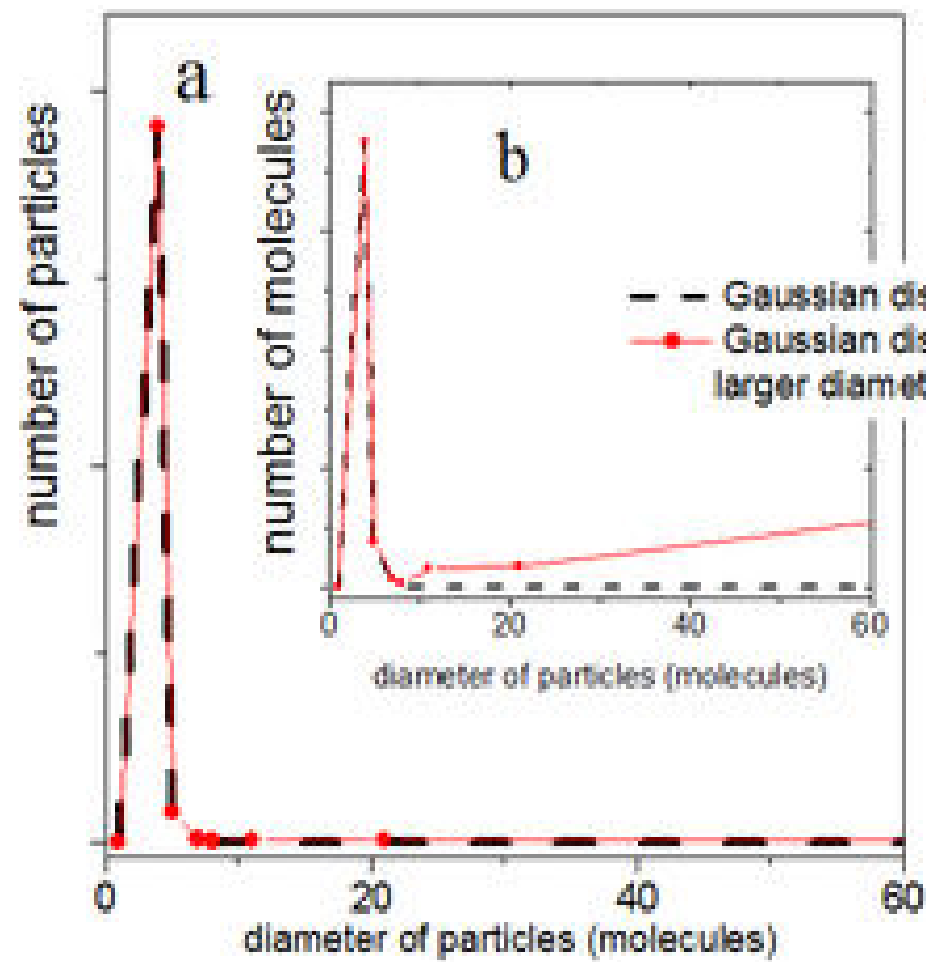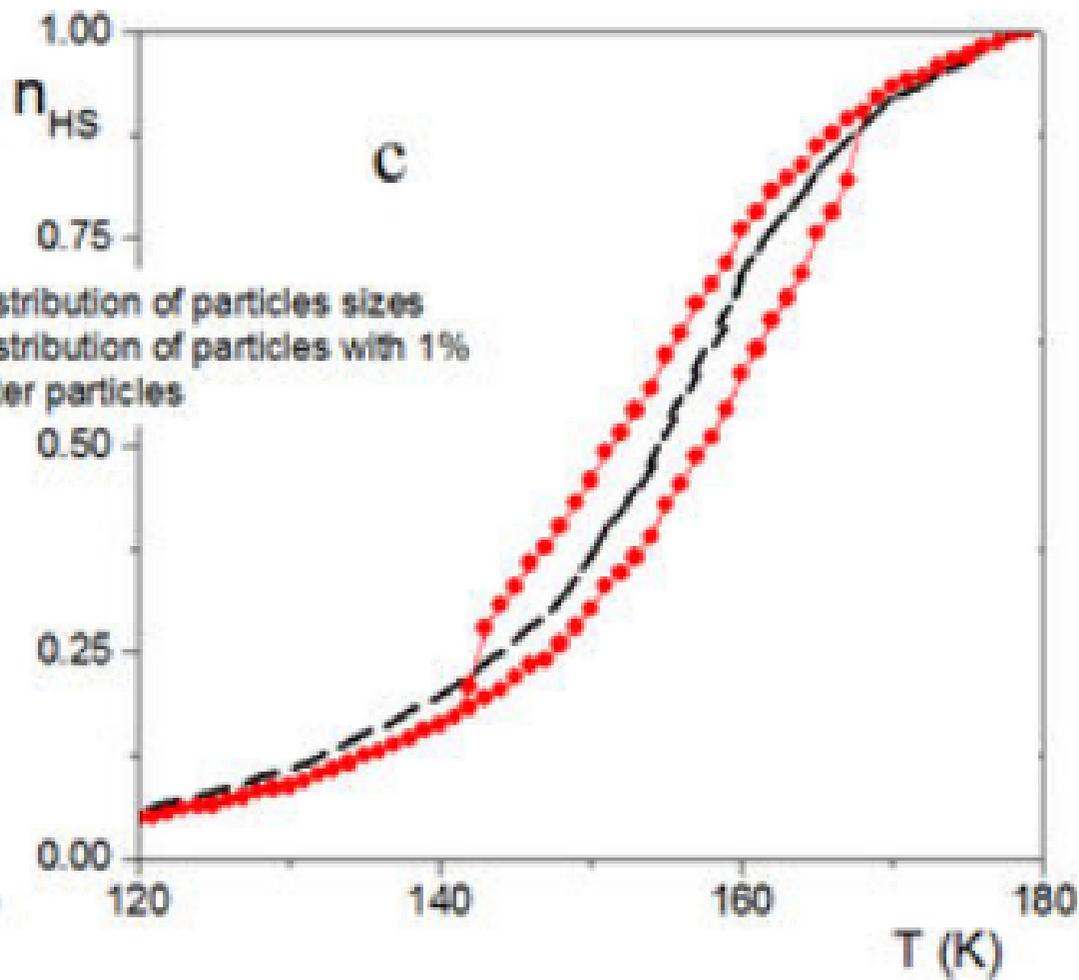

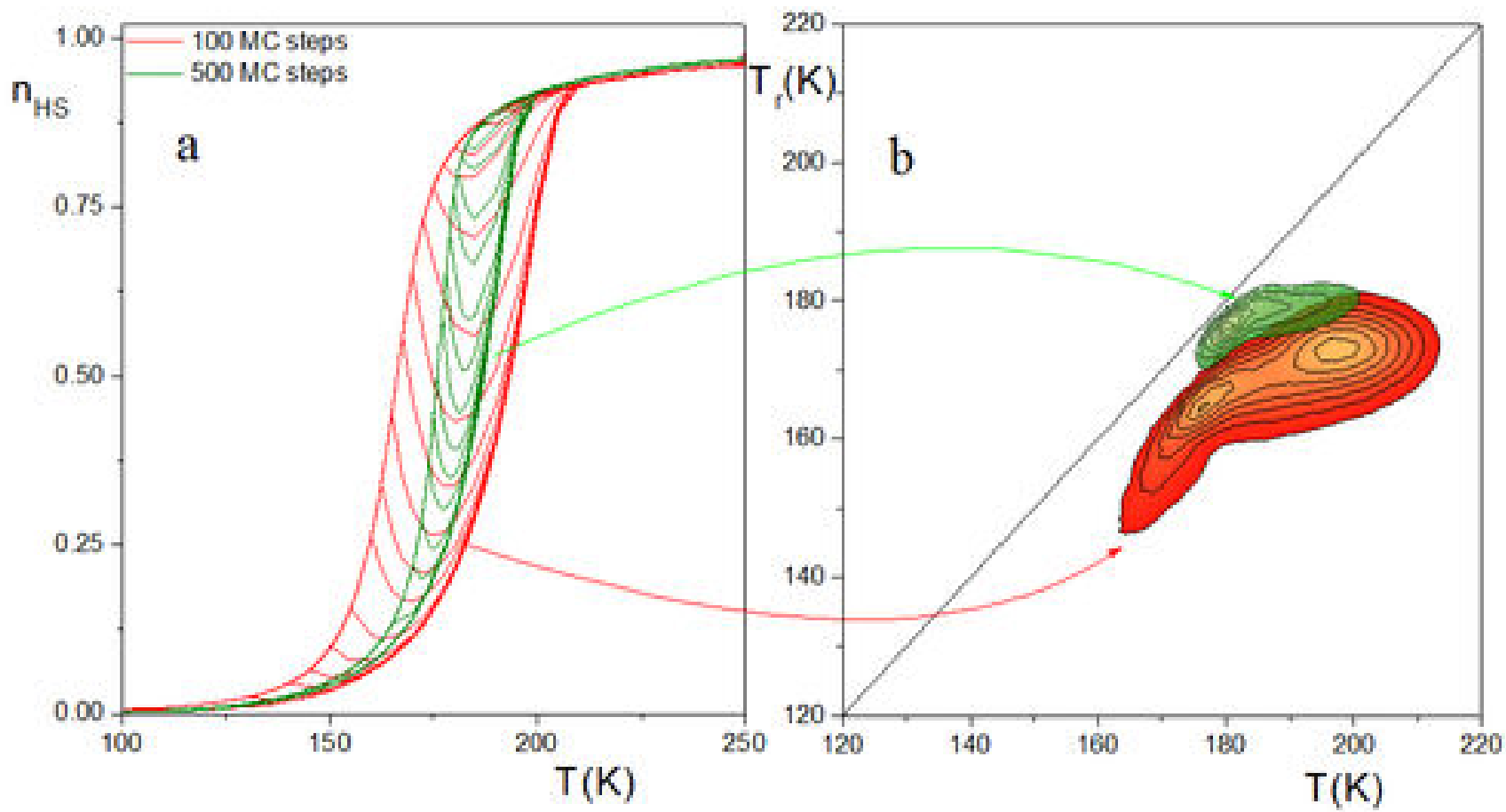

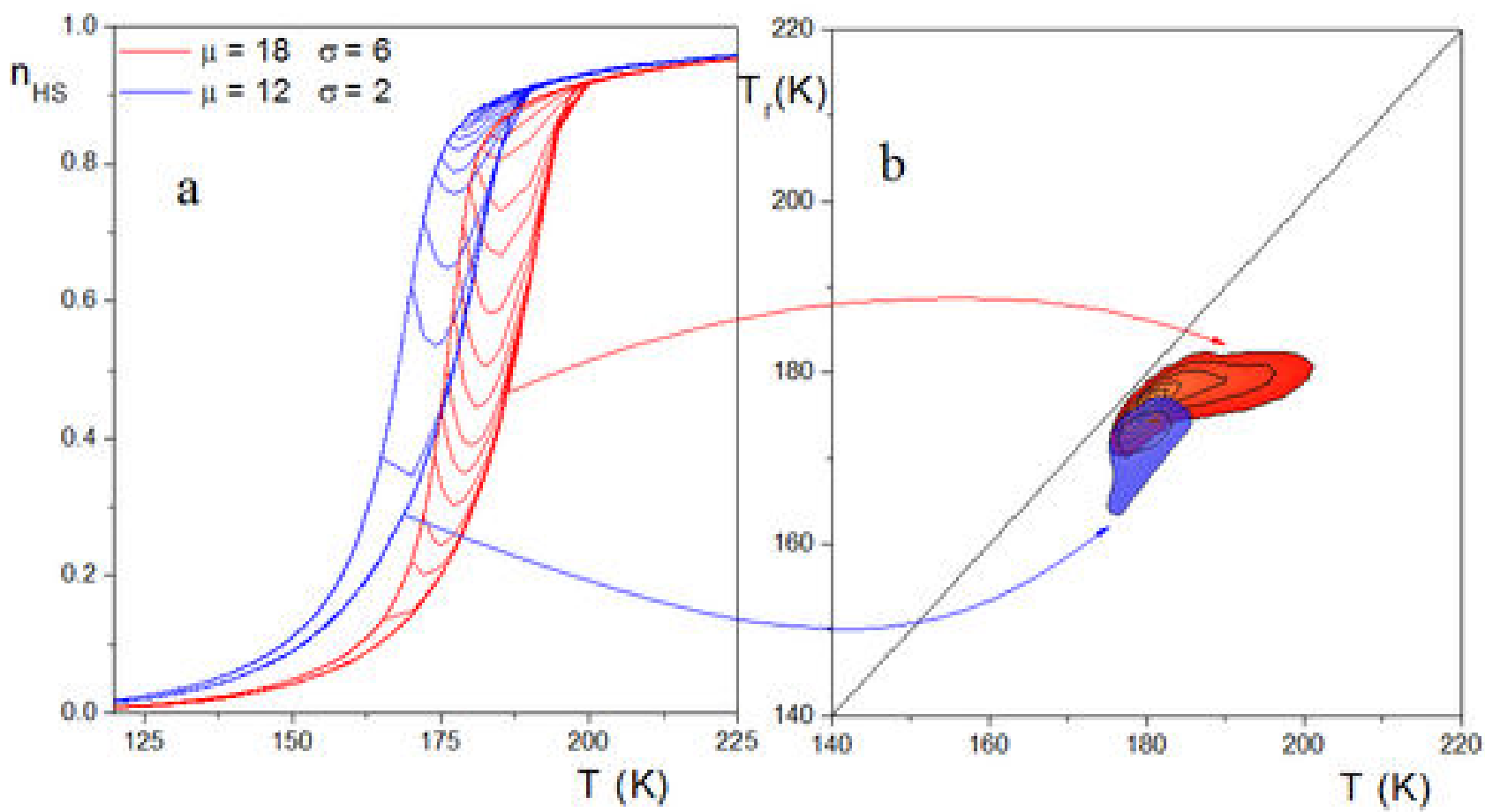